\documentclass{article}

\usepackage[nonatbib,preprint]{neurips_2021}

\usepackage{hyperref}

\usepackage[utf8]{inputenc} 
\usepackage[T1]{fontenc}    
\usepackage{url}            
\usepackage{booktabs}       
\usepackage{amsfonts}       
\usepackage{nicefrac}       
\usepackage{microtype}      
\usepackage{xcolor}         
\usepackage{soul}
\usepackage{subfigure}
\usepackage{enumitem}

\usepackage{graphicx}

\usepackage{amsmath} 
\usepackage{amssymb}

\usepackage{wrapfig}

\newcommand{\bq}{\mathbf{q}}
\newcommand{\bs}{\bf s}

\newcommand{\by}{\mathbf{y}}

\newcommand{\bZ}{\mathbf{Z}}
\newcommand{\bw}{\mathbf{w}}
\newcommand{\bW}{\mathbf{W}}

\newcommand{\bQ}{\mathbf{Q}}

\RequirePackage{fix-cm} 

\title{Quantum Annealing Formulation for\\Binary Neural Networks}

\author{%
  Michele Sasdelli\qquad Tat-Jun Chin \\
   School of Computer Science, The University of Adelaide\\
  Adelaide SA 5005, Australia \\
  \texttt{\{michele.sasdelli,tat-jun.chin\}@adelaide.edu.au} \\
}

\begin{document}

\maketitle

\begin{abstract}
  
Quantum annealing is a promising paradigm for building practical quantum computers. Compared to other approaches, quantum annealing technology has been scaled up to a larger number of qubits. On the other hand, deep learning has been profoundly successful in pushing the boundaries of AI. It is thus natural to investigate potentially game changing technologies such as quantum annealers to augment the capabilities of deep learning. In this work, we explore binary neural networks, which are lightweight yet powerful models typically intended for resource constrained devices. Departing from current training regimes for binary networks that smooth/approximate the activation functions to make the network differentiable, we devise a quadratic unconstrained binary optimization formulation for the training problem. While the problem is intractable, i.e., the cost to estimate the binary weights scales exponentially with network size, we show how the problem can be optimized directly on a quantum annealer, thereby opening up to the potential gains of quantum computing. We experimentally validated our formulation via simulation and testing on an actual quantum annealer (D-Wave Advantage), the latter to the extent allowable by the capacity of current technology.
  
\end{abstract}

\section{Introduction}

Estimates on when a practical quantum computer will be available vary from years to decades. There are also different approaches for modeling and building quantum computers, with different maturation timelines. A major paradigm is gate quantum computing (GQC), whose realization is exemplified by Google Sycamore (53 qubits) and IBM Q System One (20 qubits). Another major category is adiabatic quantum computing (AQC), which is implemented through devices called quantum annealers. Representative quantum annealers are produced by D-Wave Systems, whose latest version D-Wave Advantage contains more than 5000 qubits. While a direct comparison between different paradigms are not always meaningful since the fundamental operations are distinct, current quantum annealers have been scaled up to a larger number of qubits than quantum gate arrays.

The potential of quantum computers to monumentally shift computing, including machine learning, is not lost on physicists and machine learning researchers alike. Quantum machine learning (QML) is currently a rapidly growing field, with already several survey papers~\cite{Quantum_machine_learning_biamonte_nature,ramezani20}, monographs~\cite{quantum_machine_learning_with_python_pattanayak, Quantum_Machine_Learning_What_Quantum_Computing_Means_to_Data_Mining_Wittek} and programming libraries~\cite{tensorflow_quantum} on the topic\footnote{The other direction is also active: using machine learning to enhance quantum computing, e.g.,~\cite{mavadia2017prediction}.}.
Achievements in QML include developing quantum extensions to classical machine learning algorithms such as principal component analysis (PCA), K-nearest neighbours (K-NN), K-means clustering and decision trees~\cite{Quantum_machine_learning_biamonte_nature}. With the dominance of neural networks in machine learning, it is unsurprising that there is also significant activity in improving or extending various aspects of neural networks with quantum computing~\cite{jeswal19}.
Major directions including formulating quantum equivalents of classical neural network architectures~\cite{schuld2014quest,torrontegui2019unitary,fawaz2019training,kappen2020learning,khoshaman2018quantum}, improving the optimization (e.g., training) of neural networks with quantum computing~\cite{wu2020scrambling,beer2020training}, and even training a fully quantum neural network ``end-to-end''  on quantum data~\cite{beer2020training}.




Broadly speaking, a larger proportion of the works on QML have been framed under the GQC model, perhaps due to the broader class of problems that can be solved under the model. Comparatively fewer QML efforts have been aimed at AQC, whose realisation via quantum annealers is more suited for solving (specific types of) optimization problems. Nonetheless, with the steady increase in the capacity of quantum annealers, it is worthwhile to examine the technology for QML.

\subsection{Contributions}

In this work\footnote{The code is available at: \href{https://github.com/sasdelli/quantum_annealing_BNN}{https://github.com/sasdelli/quantum\_annealing\_BNN}}, we explore training binary neural networks (BNN) (e.g.,~\cite{fawaz2019training,bengio2013estimating,rastegari2016xnor}) using quantum annealers. BNNs are lightweight yet powerful~\cite{rastegari2016xnor} models for machine learning that are suitable for resource constrained devices. Current training regimes for BNN require smoothing or approximating the activation functions to enable first order methods such as stochastic gradient descent (SGD) on the learning of the binary (discrete) weights \cite{bengio2013estimating}. Our core contribution is to formulate the training of BNN as quadratic unconstrained binary optimization (QUBO). While the training problem remains intractable, i.e., the cost of training increases exponentially with the number of weights, our formulation is directly amenable to quantum annealers.  We experimentally validated our formulation via simulation as well as on an actual quantum annealer (the state-of-the-art D-Wave Advantage), the latter to the extent allowable by its current capacity and availability. Our results indicate that our formulation is sound and demonstrate the feasibility of quantum annealing to train BNNs.




\section{Related work}

\subsection{Quantum annealing for machine learning}

As alluded to above, both GQC and AQC have been investigated for QML.
Since our work exploits AQC for QML, we will survey only works in this area.
Interpreting linear regression as a basic form of machine learning, adiabatic quantum linear regression has been proposed~\cite{Prasanna_Date_Adiabatic_Quantum_Linear_Regression}.
In~\cite{Lorenzo_Bottarelli_A_QUBO_Model_for_Gaussian+Process_Variance_Reduction}, quantum annealing is used to minimize the posterior variance of Gaussian Processes (GP).
Neven et al.~\cite{neven2008training,neven2012qboost} and Wilsh et al.~\cite{willsch2020support} respectively formulate and solve the training of binary classifiers and support vector machines (SVM) using quantum annealing.
Another  prominent example is the quantum formulation of generative models on quantum annealers.
This can be done with Boltzmann machines, which have a natural implementation in spin lattice quantum annealers~\cite{winci2020path}, or 
the annealers could be used to generate samples for generative models~\cite{wilson2019quantum}.

Given the rapid growth of QML, our survey is likely incomplete. Nonetheless, we are arguably the first to consider AQC for BNN as well as to present empirical results on an existing quantum annealer. 

\subsection{Binary neural networks}\label{sec:bnnsurvey}

Ther are several implementations of BNNs~\cite{qin2020binary}. Generally speaking, they are neural networks with weights and/or biases and/or activations constrained to binary (typically ${-1,1}$).
The goal of the approach is to produce resource efficient models that use primarily binary operations, such as XNOR operations and bitcount operations~\cite{conti2018xnor}.
As deep learning model become better, they are also growing in computational resources requirement~\cite{openai_compute}, both in training and inference.
BNNs aim to reduce the computational requirements during inference.
However, the standard training procedures used to train BNNs require ``float'' number operations on a continuous version of the BNN to make it amenable to SGD optimization.
``straight-through'', the most common heuristic uses a mix of ``binarized'' forward pass and  continuous backpropagation on approximated continuous weights and smoothed activation functions\cite{bengio2013estimating}.
Hence, the training can only minimally benefit in performances from the binary structure of the trained network.
A direct optimization of the binary problem, although preferable, cannot be done with first order method without the use of similar approximations.
From a hardware perspective, binary neural networks are an ideal type of model by using bit wise operations that are natively implemented in hardware~\cite{conti2018xnor}. 





\section{Brief introduction to quantum annealing}\label{sec:aqcintro}

Quantum annealers are the practical realization of AQC. At its core, a quantum annealer solves optimization problems through energy minimization of a physical system.
A Hamiltonian defines the energy profile of a quantum system, which is composed of a number of interacting qubits.
In a quantum annealer, the Hamiltonian is the sum of the \emph{initial} Hamiltonion and \emph{final} Hamiltonion---the target optimization problem is encoded in the latter. The system is initialized in the ground state (lowest energy configuration) of the initial Hamiltonion, which is then evolved (annealed) such that the final Hamiltonian dominates in the overall system. Assuming adiabaticity (the evolution is slow enough) and negligible thermal noise, the final state of the system (the desired solution) will be in the ground state of the final Hamiltonion. See~\cite[Chap.~8]{scherer2019mathematics} for more details on quantum annealing.

The Ising model allows to calculate the energy of the Hamiltonian
\begin{align}\label{eq:bin_quad}
    f(\bq) =\sum_{n} Q_{nn}q_n  + \sum_{n < m} Q_{nm} q_n q_m =  \bq^T \bQ \bq
\end{align}
corresponding to an $N$-qubit quantum state $\bq = \left[q_1, q_2, \dots, q_N\right]$, where each $q_n \in \{0,1\}$, $\bQ \in \mathbb{R}^{N \times N}$ and $Q_{nm}$ is the element at the $n$-th row and $m$-th column of $\bQ$. In a quantum annealer, the target problem is specified as an instance of the energy  minimization problem
\begin{align}\label{eq:generic_qubo}
    \min_{\bq \in \{0,1\}^N} \bq^T\bQ\bq,
\end{align}
which is a QUBO.
The combinatorial nature of QUBO renders it intractable on a classical machine. A quantum annealer, by virtue of the physical processes described above which allows $\bq$ to evolve through superposed states (a phenomenon called ``quantum tunneling''), may solve such a problem efficiently; see~\cite[Chapter~8]{scherer2019mathematics}.


\subsection{Quantum processing unit}\label{sec:qpu}

The quantum processing unit (QPU) of a quantum annealer essentially contains an Ising system of interconnected qubits, where the strength of the interconnections can be modified/programmed to define the quadratic function $f(\bq)$. In D-Wave quantum annealers, the QPU is implemented as a lattice (i.e. grid) of superconducting flux qubits~\cite{harris2010experimental}.
The interaction between qubits are provided by couplers, that allow the physical interaction between the qubits. The strengths of these couplers are what we can ``program''.
The geometry of the lattice limits the possible available couplings, ultimately defining the computation graph.



\begin{figure}[ht]\centering
  \subfigure[]{\includegraphics[width=0.4\textwidth]{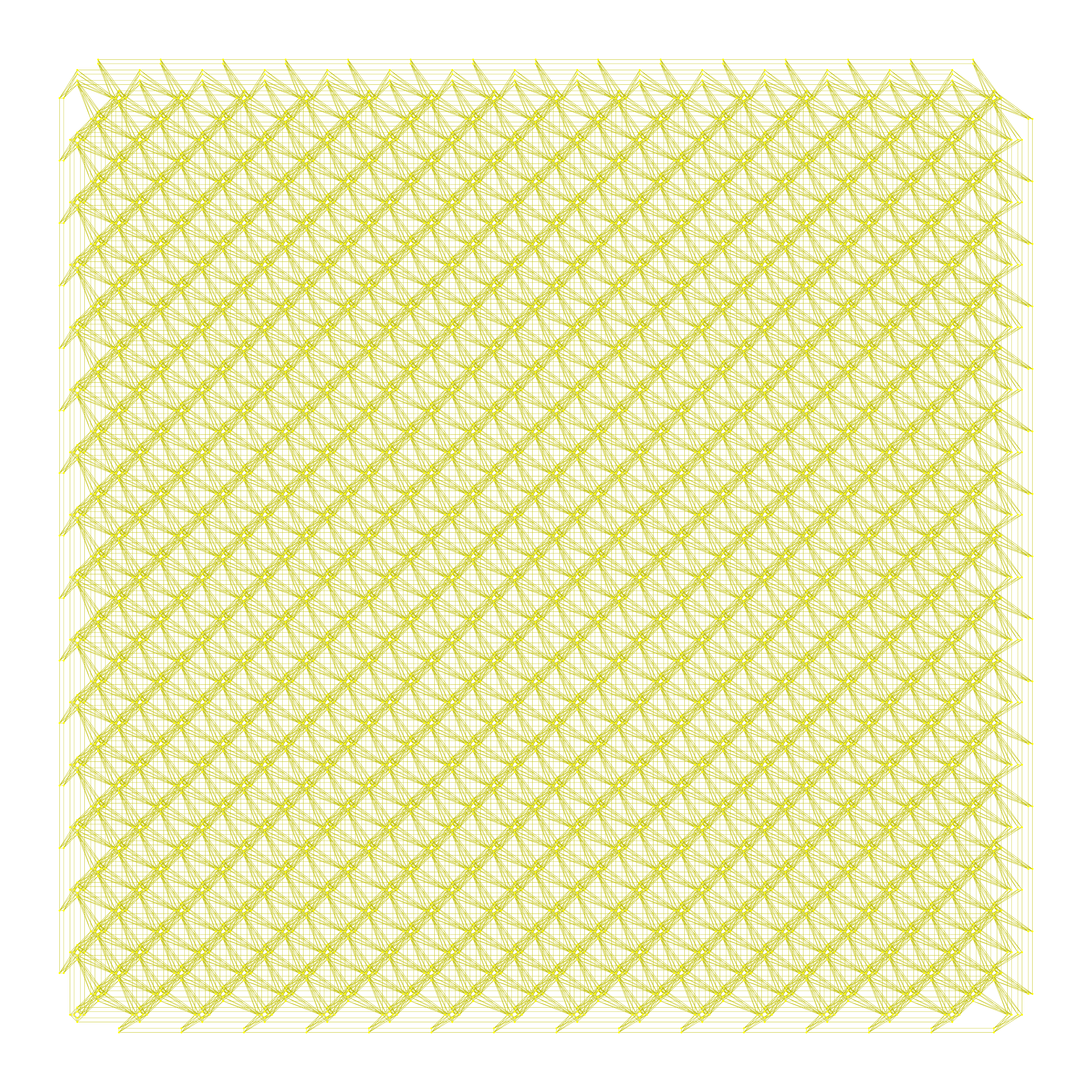}}
  \subfigure[]{\includegraphics[width=0.4\textwidth]{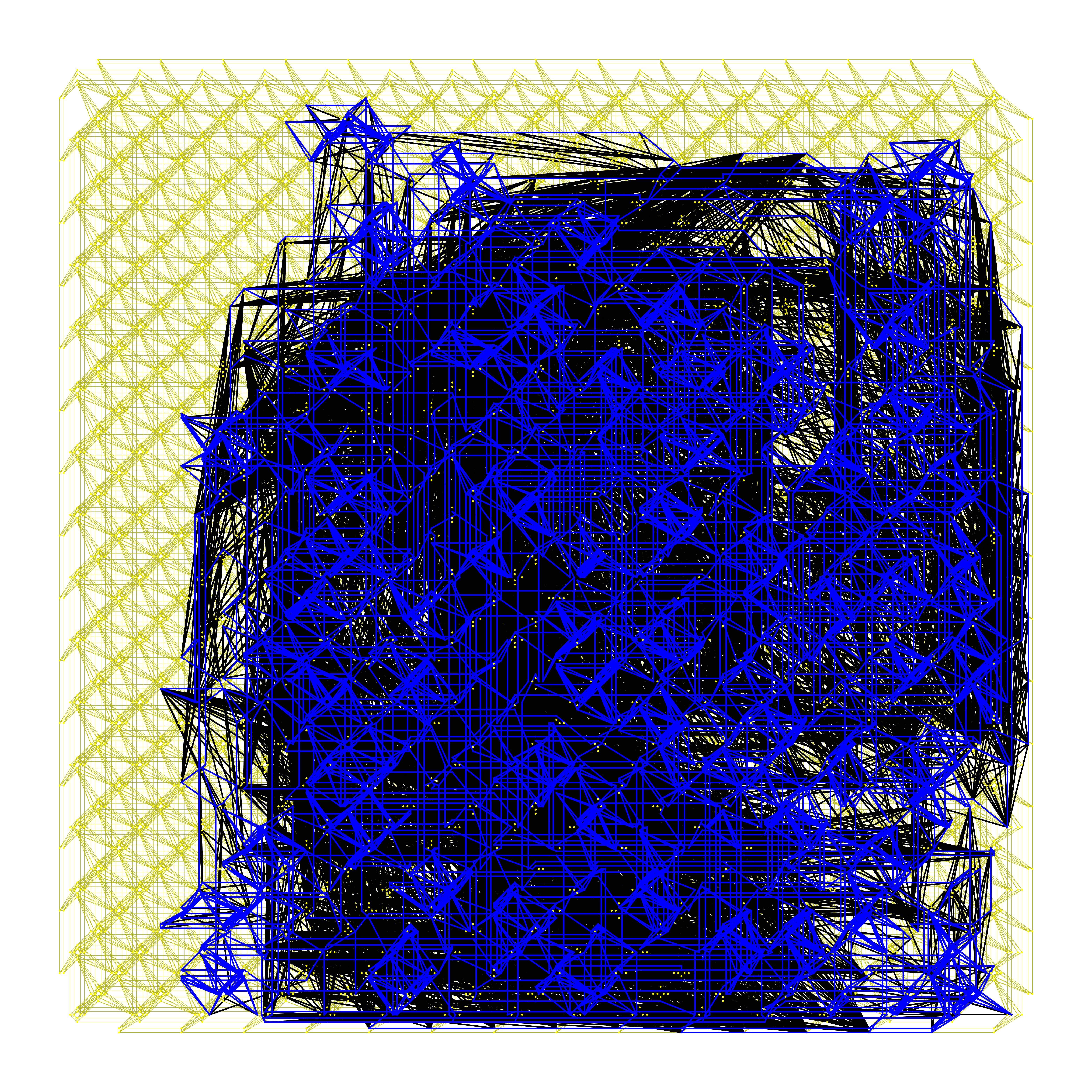}}  
  \caption{(a) Topology of the Pegasus QPU of D-Wave Advantage, where the yellow lines indicate qubit interconnections. See D-Wave documentation~\cite{dwave_pegasus} for a clearer illustration. (b) The embedding of an instance of the BNN training problem using our formulation (Sec.~\ref{sec:bnnqubo}) is illustrated by blue lines (chain breaks in black). This embedding, which employs $\approx 3000$ qubits (a large part of the capacity), corresponds to a network size of $7$ inputs,  $7$ hidden neurons, $2$ layers and dataset size of $8$.
  }
  \label{fig:embedding}
\end{figure}

Due to chip geometry limitations, however, the interconnections between qubits in a practical QPU is usually incomplete (not all quadratic terms are available) and the precision of $\bQ$ is lower than that of a classical computer. This limits the density and structure of the QUBO that can be solved.
In the case of D-Wave Advantage whose QPU implements the Pegasus architecture (around $5000$ qubits), each qubit is connected to $15$ other qubits; see Fig.~\ref{fig:embedding} for the topology.
Given a desired $\bQ$, an embedding step~\cite{cai2014practical} needs to be performed to find the equivalent QUBO that is feasible on the QPU. Typically, this will involve increasing the dimensionality of the problem.

\section{QUBO formulation for BNN}

In this section, we develop a QUBO formulation for the BNN training problem, so as to facilitate the training of BNNs on a quantum annealer.

\subsection{BNN}\label{sec:bnn}

Generally speaking a BNN is a series of layers $\ell \in \{0,1,\dots,L-1 \}$  with binary weights ${\bW}^\ell$ and inputs $\by^\ell$ at each layer. Using $\{-1,1\}$ (up and down ``spins states'') to specify the weights and inputs~\cite{courbariaux2016binarized}, the output at layer $\ell+1$ is
\begin{align}\label{eq:layer}
{\bf y^{\ell+1}}=\mathsf{sgn}({{\bf W}^{\ell} \by^{\ell}}),
\end{align}
and the overall output is the output of the final layer which contains the single neuron
\begin{align}\label{eq:bnn}
y^{L} = \mathsf{sgn}({\bf W}^{L-1}\mathsf{sgn}({\bf W}^{L-2} ... \; \mathsf{sgn}({\bf W}^0{\bf y}^0) ... )) = f_{BNN}({\bf w})({\bf y}^0).
\end{align}
Without loss of generality, we do not include the bias terms in the BNN (this is also aligned with~\cite{rastegari2016xnor,bulat2019xnor}); in any case, introducing binary biases is mathematically straightforward but will render our notation more complicated. The nonlinearity in the BNN is implemented by the sign function
\begin{align}
    \mathsf{sgn}(x)=\frac{x}{|x|},
\end{align}
where $x \in \mathbb{Z}$. For further background (motivation, practical usage, etc.) of BNNs, see Sec.~\ref{sec:bnnsurvey}.



\subsection{BNN training problem}

Given an input and target output pair $({\bf{y^0}}, \hat{y})$ from the training dataset, let $L( \hat{y} , {y^L})$ evaluate the loss for the training datum, where $y^L$ is the network output corresponding to $\by^0$ as defined in~\eqref{eq:bnn}. Minimizing the aggregate loss for the dataset over the binary weights $\bW^0, \bW^1, \dots, \bW^{L-1}$ is a discrete optimization problem. In particular, for the $0-1$ indicator loss together with regularization terms to improve generalizability, Icarte et al.~\cite{icarte2019training} showed that the optimal BNN weights can be learned via a mixed integer program (MIP). In general, BNN training is intractable~\cite{Goodfellow-et-al-2016}, i.e., the cost of searching for the optimal weights scales exponentially with the number of weights. In previous works, smoothing or relaxation are required to enable gradient-based methods (see Sec.~\ref{sec:bnnsurvey}).






\subsection{BNN training as QUBO}\label{sec:bnnqubo}

Let $W^\ell_{ji}$ be the element at the $j$-th row and $i$-th column of $\bW^\ell$, and $y^\ell_i$ as the $i$-th element of $\by^\ell$. Let
\begin{align}
\bZ^{\ell+1} = \bW^{\ell}\by^{\ell},
\end{align}
where the element at the $j$-th row and $i$-th column of $\bZ^{\ell+1}$ is 
\begin{equation}
{ Z}_{ij}^{\ell+1}= { W}^\ell_{ji}y_{i}^\ell,
\label{perceptron}
\end{equation}
i.e., $\bZ^{\ell+1}$ contains the element-wise multiplication of the weights and inputs from the previous layer. Expanding~\eqref{eq:layer}, the output of the $j$-th neuron at layer $\ell+1$ is
\begin{equation}
y_j^{\ell+1} = \mathsf{sgn}\left( \sum_{i} {Z}_{ij}^{\ell+1} \right).
\label{nonlin}
\end{equation}
Recall that the derivations above are based on taking the weights and inputs as spin variables. 

Given a training datum $({\bf{y^0}}, \hat{y})$, the 0-1 loss is defined as
\begin{align}\label{eq:01lost}
L_{0-1}( \hat{y} , {y^L}) = \begin{cases}
0 & \hat{y} = y^L;\\
1 & \text{otherwise}.
\end{cases}
\end{align}
The corresponding training problem can be formulated as the quadratically constrained binary optimization (QCBO) problem
%
\begin{align}\label{eq:qcbo}
  \min_{ \{\bW^\ell,  \bZ^\ell\}^{L-1}_{\ell = 0},
  \{ \by^\ell\}^{L-1}_{\ell = 1},
  y^L }
\quad & \left(\frac{y^L - \hat{y}}{2}\right)^2   \\
\textrm{s.t.} \quad & y^L \in \{-1,1\},\\
& W_{ij}^\ell, y_i^\ell, Z_{ij}^\ell \in \{-1,1\},\\
& Z_{ij}^\ell = W_{ji}^\ell y_i^\ell, \\
& \left( \sum_i Z_{ij}^\ell \right) y_j^\ell \ge 0.
\label{eq:qcbo_constr}
\end{align}
where constraint~\ref{eq:qcbo_constr} is equivalent to constraint~\ref{nonlin} because the product of two numbers that agree in sign is larger or equal than zero.
Note that squaring the 0-1 loss does not qualitative change the problem. To simplify subsequent notation, we collate all the binary variables into the $N$-element spin vector
\begin{align}
{\bf s} =
\left( y_0^1, y_1^1, ... \, Z_{0,0}^0, Z_{0,1}^0 ... \, Z_{0,0}^1, Z_{0,1}^1, ... \, W_{0,0}^0, W_{0,1}^0, ... \, y^L \right)
\end{align}
and rewrite the QCBO above as
\begin{align}\label{eq:qcbo2}
\begin{aligned}
\min_{ {\bf s} } \quad & \left( \frac{ s_{\textrm{last}} -\hat{y} }{2}\right)^2\\
\textrm{s.t.} \quad & \text{constraints\, from\, QCBO},
\end{aligned}
\end{align}
where $s_{\textrm{last}}$ is the final element of $\bs$.


\paragraph{Penalty method}

We now convert the QCBO into a QUBO by incorporating the constraints directly into the quadratic objective at the expense of additional $\bf s$ variables. First, our spin variables ${\bf s} \in \{-1,1\}^{{N}}$ can be converted into the equivalent binary variables
\begin{align}
{\bf q } = \frac{({\bf s}+1)}{2} \in \{0,1\}^N.
\end{align}

The equality~\eqref{perceptron} can be rewritten using binary variables as
\begin{align}\label{eq:equality}
q_{Z^{\ell+1}_{ij}} = \mathsf{XNOR}( q_{{W}^\ell_{ji}}, q_{y^\ell_i}),
\end{align}
where hereafter the subscript $s$ in $q_s$ indexes the binary variable in $\bq$ corresponding to the spin variable $s$. Now, $\mathsf{XNOR}$ is the logical gate
\begin{align}
    1 = \mathsf{XNOR}(0,0), \;\;\;\; 0 = \mathsf{XNOR}(0,1), \;\;\;\; 0 = \mathsf{XNOR}(1,0), \;\;\;\; 1 = \mathsf{XNOR}(1,1).
\end{align}
Introducing a penalty parameter $P \in \mathbb{R}^+$, violations to the logical statement $q_3=\mathsf{XNOR}(q_1,q_2)$ can be penalized via the cubic polynomial
\begin{align}\label{eq:cubic}
P(1-q_1-q_2-q_3+2q_1q_2+2q_2q_3+2q_1q_3-4q_1q_2q_3).
\end{align}
Following \cite{glover2018tutorial}, the cubic term $q_1q_2q_3$ can be reduced to a quadratic term with an additional ancillary binary variable $b$, i.e.,
\begin{align}
q_1q_2q_3 \approx b q_3 + P( 3b + q_1 q_2 -2 q_1 b -2 q_2 b ).
\end{align}
The penalty term~\eqref{eq:cubic} thus also reduces to the quadratic (in the bit variables) polynomial
\begin{align}\label{eq:g}
\begin{aligned}
    g(q_1, q_2, q_3, b) = &P(1-q_1-q_2-q_3+2q_1q_2+2q_2q_3+2q_1q_3\\ &-4(  b q_3 + P( 3b + q_1 q_2 -2 q_1 b -2 q_2 b ) ))
    \end{aligned}.
\end{align}

%
%

Moving on to the nonlinearity of the sum (Eq.~\ref{nonlin}), which becomes Eq.~\ref{eq:qcbo_constr} as a spin constraint.
Let us first look at the case with a sum with three terms (three neurons).
The sum of the three spins can be: $\{-3, -1, 1, 3\}$.
The corresponding sum of three bits can be $\{0, 1, 2, 3\}$.
If we write it in binary notation, it is $\{00,01,10,11\}$. 
We can see that the required output of Eq.~\ref{nonlin} corresponds to the highest bit in binary notation.
the penalty term for for the activation of sum requires one new auxiliary bit $a$ for the units of the binary notation.
The output bit $q^{(out)}$ is the bit representing $2$ in the binary notation.
The number is then: $ 2q^{(out)} + a $.
The constraint that we want to enforce can be rewritten as: $ a + 2q^{(out)} = q_0^{(in)} + q_1^{(in)} + q_2^{(in)}$.
With $P$ a large number working as a penalty, we can build a the quadratic term $h$: 
\begin{align}
h(a, q^{(out)} , q_{0}^{(in)} , q_{1}^{(in)}, q_{2}^{(in)} )=P(-a -2q^{(out)} + q_{0}^{(in)} + q_{1}^{(in)}+ q_{2}^{(in)}  )^2
\label{eq:h}
\end{align}
%
%
%
In the general case, if the number of inputs of a layer is restricted to dimension $2^{n}-1$, and $a_{k}$ are auxiliary bits. 
The penalty term for the activation of the sum becomes:
\begin{align}
h(a_{0}, ..., a_{n-2},q^{(out)}, q_{0}^{(in)}, ..., q_{2^{n}-1}^{(in)}  )=P\left(-\sum_{i=0}^{n-2} 2^i a_{i}   -2^{n-1}q^{(out)} + \sum_{k=0}^{2^n-1} q_{k}^{(in)}  \right)^2.
\label{eq:constr_general}
\end{align}
We are basically ``counting'' how many $Z$ bits are ones using the ancilla bits and the output bit, representing the number in base two, and using the most significant bit $q^{(out)}$ as output.

The term $h$~\eqref{eq:constr_general} is quadratic in the bits, and hence can be added to the quadratic loss~\eqref{eq:bin_quad} for each neuron in each layer.
We need to ``extend'' our spin vector $\bf s$ and corresponding binary vector $\bf q$ to include the ancillary variables:
\[ {\bf s} = \left( y_0^1, y_1^1, ... Z_{0,0}^0 ... W_{0,0}^0 ... b_{0,0}^1 ... a_{0,0}^0, a_{0,1}^0 ...  \right) \]

Changing the constraints in problem~\eqref{eq:qcbo} to their penalty forms, we arrive at the QUBO:
\begin{align}\label{eq:qubo}
\begin{aligned}
\min_{ \bf q \in { \{0,1\}^M }} \quad & \left({ q_{y^L} -   \hat{b}}\right)^2 + \sum_{\ell,i,j} g( q_{Z^{\ell+1}_{i,j}}, q_{W^\ell_{i,j}}, q_{y^\ell_{i}}, b^\ell_{i,j})  + \sum_{\ell,i,j} k( a_{i,j}^\ell, ..., q_{y_j^{\ell+1}}, q_{Z_{k,j}^\ell} ),  \\
\end{aligned}
\end{align}
which has the exact same form as~\eqref{eq:generic_qubo}. The first term is the squared $0-1$ indicator loss with the label $\hat{b} = (\hat{y}+1)/2$, the second term enforces the element-wise muliplications~\eqref{perceptron} or, equivalently, the XNORs~\eqref{eq:equality}, the last term enforces the summations and nonlinearities~\eqref{nonlin}.
For inputs require a larger number of auxiliary bits. The number of ancilla bits required by the nonlinearities scales only logarithmically with the input size of the layer ($\log_2({N})$). The ancilla bits required by the element-wise multiplications~\eqref{perceptron} due to the cubic terms  above are equal to the number of the weights.


\paragraph{Optimization over multiple data points}

Our derivations above are based on minimizing the training error of a single training datum. Given a dataset $\{(\by^0_d, \hat{y}_d )\}^{D}_{d=1}$, additional binary variables of the type $q_{y_i^\ell}$, $q_{Z_{ji}^\ell}$, $b_{k,j}^\ell$ and $a_{k,j}^\ell$ will need to be introduced to account for the other data points; however, the weight variables $q_{W_{ij}^\ell}$ are common. In Sec.~\ref{sec:experiments}, we will report results on training on multiple data points. To reduce clutter, we refer the reader to the supplementary material for detailed derivations of the QUBO for training a BNN on a dataset $\{(\by^0_d, \hat{y}_d )\}^{D}_{d=1}$.

\paragraph{One-shot training versus online training}

The training problem described thus far concerns ``one-shot'' training, i.e., estimating the weights in a single pass over all the training data, which is analogous to the training of a support vector machine (SVM). To scale our BNN training to large datasets, it is vital to perform online or incremental training over multiple batches/epochs of the training dataset. In Sec.~\ref{sec:discussions}, we will describe initial ideas for online training of BNNs on quantum annealers.

\subsection{Embedding QUBO in QPU}\label{sec:embedding}

To embed our training problem~\eqref{eq:qubo} onto the Pegasus topology on D-Wave Advantage, we used the $\mathsf{minorminer}$ tool from D-Wave~\cite{cai2014practical}. Conceptually, the embedding algorithm finds the mathematically equivalent QUBO that fits on the QPU, and this usually increases the dimensionality of the problem. Although finding the embedding is an optimization by itself, the cost can be ignored since in the BNN training problem, the structure of the QUBO is fixed for a given BNN architecture.

The topology of the graph of the non-zero quadratic terms of our QUBO will not, in general, be a subgraph of the Pegasus graph. An obvious example is when bits have "too many" connections. The embedding procedure replaces some of the bits of the original QUBO formulation with a "chain" of bits, coupled with a strong coupling (using additional penalties), distributing the connections of the original bit between the elements of the chain.

Fig.~\ref{fig:embedding} illustrates the embedded QUBO for an instance of the BNN training problem~\eqref{eq:qubo} involving $7$ inputs, $7$ hidden neurons and $2$ layers in the BNN and a batch size of $D = 8$. While this is a tiny network and batch size in comparison to practical neural networks, the QPU is almost fully utilised by the embedding (requires $>3000$ qubits). For reference, the corresponding QCBO and QUBO formulations respectively have $584$ and $760$ variables.

\section{Experiments}\label{sec:experiments}



We investigated a binary classification problem with the UCI ``adult'' dataset~\cite{platt1998sequential}, which contained $123$ binary attributes, $1,605$ labeled training data points and $30,956$ testing data points. As alluded in Sec.~\ref{sec:embedding}, the Pegasus archictecture was able to support only small instances of the BNN training problem, hence, in our experiments, we used only a small subset of the data to generate QUBO instances that were feasible on the Pegasus QPU while still representative of our formulation. In addition, our limited subscription to D-Wave Advantage compelled us to focus on comparing the computational performance of solving the BNN training problem. We leave large-scale evaluation of QPU-trained BNNs, including online training, as future work (as we will discuss in Sec.~\ref{sec:discussions}).

\subsection{Classical algorithms versus quantum annealing}\label{sec:exp0}

By sampling the UCI adult dataset, we generated training datasets $\{(\by^0_d, \hat{y}_d )\}^{D}_{d=1}$ as follows:
\begin{itemize}[leftmargin=1em,topsep=0em,itemsep=0.25em,parsep=0em]
\item $20$ datasets with $3$ input attributes and $D = 4$ data points each (Type A).
\item $20$ datasets with $3$ input attributes and $D = 8$ data points each (Type B).
\end{itemize}
We used a two-layer BNN with 3 input neurons and 3 hidden neurons, and 1 output neuron for Type A (resp.~Type B) instances. For each dataset and the corresponding network architecture, we defined BNN training problems in QCBO~\eqref{eq:qcbo} and QUBO~\eqref{eq:qubo} forms. In addition, the Pegasus embedding (henceforth called eQUBO; see Sec.~\ref{sec:embedding}) of the QUBO instances were also constructed. For all the QUBO and eQUBO instances, the penalty value was set to $P = 50$.

The following algorithms were executed:
\begin{itemize}[leftmargin=1em,topsep=0em,itemsep=0.25em,parsep=0em]
    \item Branch-and-bound (BnB) (as implemented on Gurobi~\cite{gurobi}) for QCBO, QUBO and eQUBO.
    \item Simulated annealing (SA) (as implemented on~\cite{qubovert2021}) for the QUBO and eQUBO. The number of iterations of SA was fixed at $1000$.
    \item Quantum annealing (QA) (as executed on D-Wave Advantage) for the eQUBO. We used default settings for the annealing schedule ($20 \mu s$) and $500$ samples (see~\cite{DWave_params}). 
\end{itemize}

Let $\bw^\ast_{BnB.X}$, $\bw^\ast_{SA.X}$ and $\bw^\ast_{QA.X}$ be the optimal binary weights returned respectively by BnB, SA and QA for formulation $X$. Since BnB conducts exhaustive research, it always finds the globally optimal weights regardless of the formulation. We thus took the BnB optimized weights (for QCBO) as the ``ground truth'', and computed the distance of another solution $\bw$ to the ground truth as
\begin{align}
 d(\bw) = \sum_{d=1}^D  |f_{BNN}(\bw^\ast_{BnB.QCBO})({\bf y}^0_d) - { \hat{y}_d}| -| f_{BNN}(\bw)({\bf y}^0_d) - {\hat{y}_d} |
\end{align}
where $f_{BNN}(\bw)(\by^0)$ is the forward mapping of the BNN~\eqref{eq:bnn} corresponding to network weights $\bw$. 

\begin{figure}
  \centering
  \includegraphics[width=1.0\textwidth]{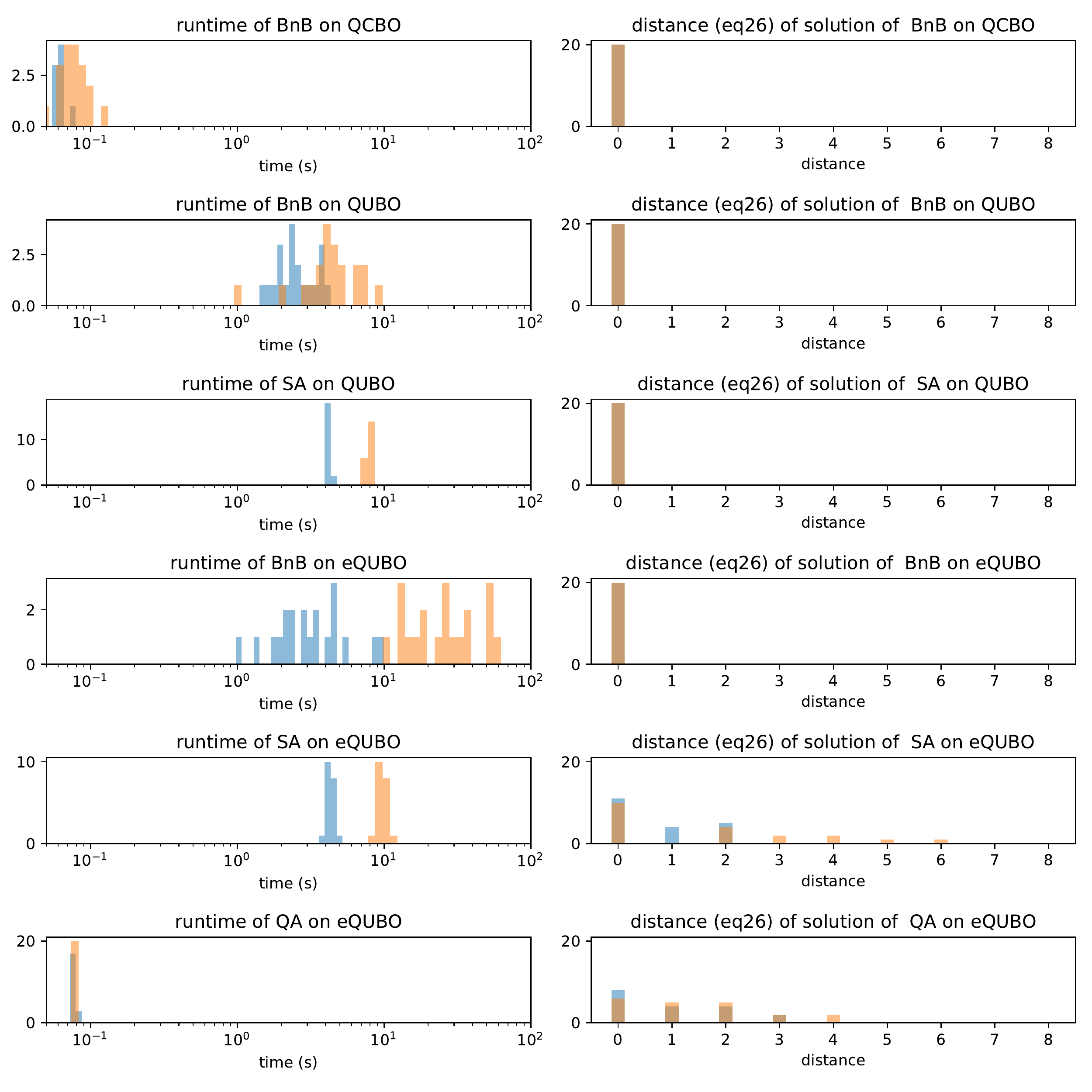}
  \caption{Histograms of runtime (in logarithmic scale) and distance of solution of different algorithms on two-layer BNNs for Type A (blue histograms) and Type B (orange histograms) datasets (see text in Sec.~\ref{sec:exp0} for details). The time reported for QA is the QPU execution time (also called access time).}
  \label{fig:results}
\end{figure}

Fig.~\ref{fig:results} illustrates the results of the experiment. In terms of runtime, BnB terminated in about $100$ ms for the QCBO instances; however, on the QUBO and eQUBO instances the runtime of BnB visibly increased (up to tens or hundreds of seconds), due to the larger number of variables it those formulations (Sec.~\ref{sec:scaling} will further examine runtime scaling). The runtime of SA on eQUBO slightly increased from that of QUBO, but for both formulations the runtime was still in the range of tens of seconds. QA terminated in about $100$ ms for all instances, due to using the same annealing schedule and fixed number of samples.

In terms of distance to the ground truth, the fact that BnB and SA on the QUBO instances yielded distances of zero indicates that our QUBO formulation~\eqref{eq:qubo} is correct. The solutions of BnB on eQUBO also returned zero distances, thus validating the embedding tool~\cite{cai2014practical}. However, the solutions of SA and QA on eQUBO did not have distances of zero uniformly; for SA, this was likely due to using fewer than optimal number of iterations. For QA, we suspect that undesirable thermal noise as a factor in the non-zero distances. However, the fact that the solutions from SA and QA on eQUBO had similar distances to the ground truth indicates a similar quality between the classical and quantum counterparts, though with vast different runtimes.

\subsection{Runtime scaling}\label{sec:scaling}

%
%
\begin{figure}
\begin{center}
  \includegraphics[width=0.8\textwidth]{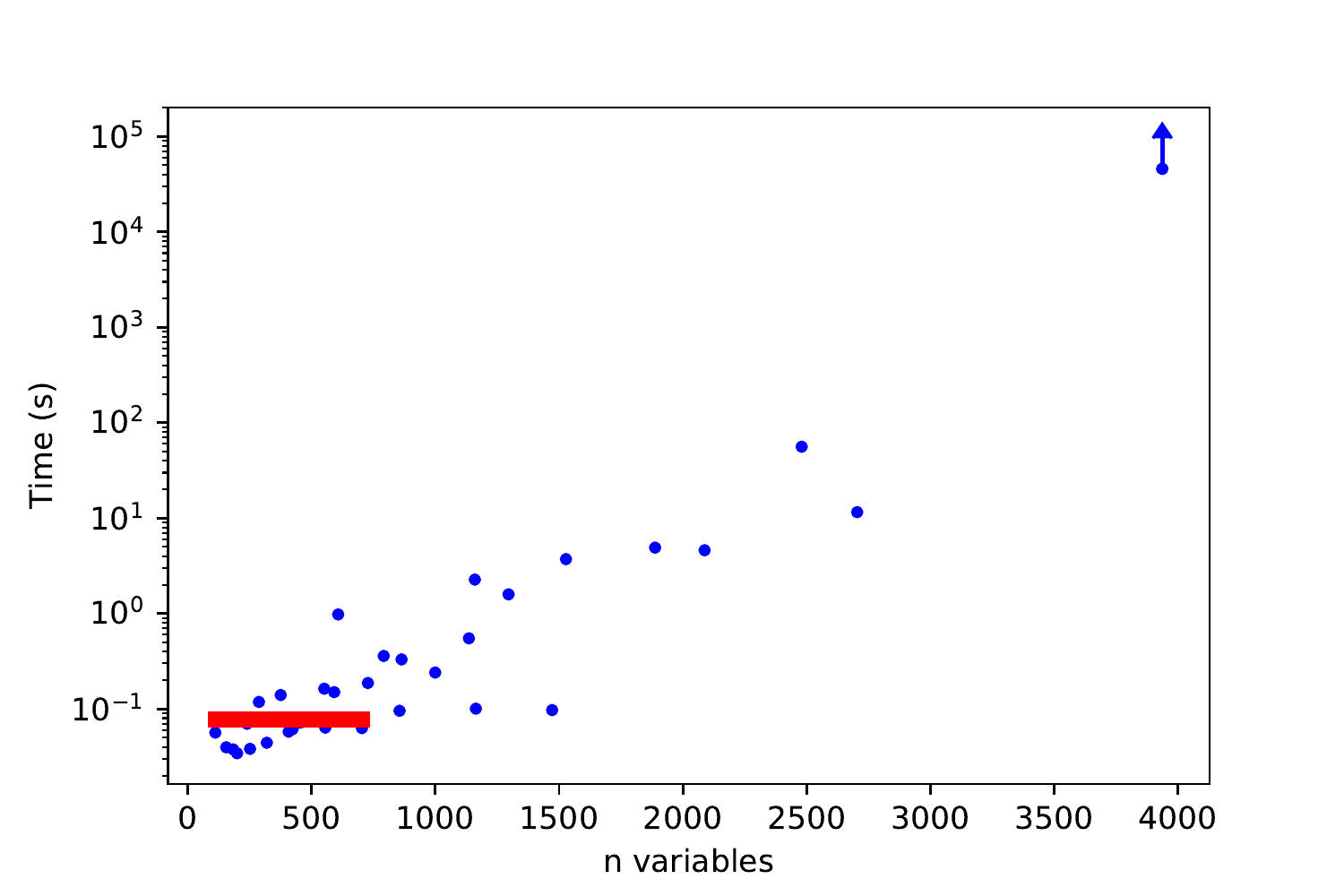}
\end{center}
  \caption{Runtime of BnB as a function of number of variables of QCBO on several BNNs. The red line represents the expected runtime on the QPU for problem sizes that can be embedded on Pegasus.}
  \label{fig:exp_time}
\end{figure}

Fig.~\ref{fig:exp_time} shows the runtime of BnB for solving QCBO~\eqref{eq:qcbo} with two-layer BNNs on several datasets of sizes $D \in [4, 32]$, input dimension between $3$ and $7$, number of hidden heurons between $3$ and $15$. As expected, the BnB runtime scales exponentially with the number of variables (blue dots). The expected runtime (red line) of the QPU for  QUBOs small enough to be embedded on Pegasus was also plotted for reference. Note that the number of qubits of quantum annealers has been roughly doubling every year.
\section{Discussions and future work}\label{sec:discussions}

\paragraph{Optimizing formulations}

On classical computers, exploiting the inherent structures (e.g., symmetries, redundancies) in optimization problems are crucial to attain speed-ups. Similarly, the performance of a quantum algorithm depends on the problem encoding. However, formulations that are optimal for classical algorithms may not be optimal for the quantum counterpart. In this paper, we focused on taking the first step---how to formulate a representative ML problem for a quantum annealer---and we leave optimizing the quantum formulation as future work. 

\paragraph{Online training}

Our training problem in Sec.~\ref{sec:bnnqubo} conducts ``one-shot'' training, i.e., estimating the weights in a single pass over all the training data, which is analogous to the training of a support vector machine (SVM). Here, we briefly describe how an online (batch-by-batch) training algorithm can be fashioned using our QUBO training problem as a subroutine. Basically, the result of the training over the past batches need to influence the QUBO of the current batch. For example, if there are multiple optimal solutions for the current batch of data, and one of these solutions happens to be the solution of a previous batch, we want it to be favored over the other ones. A learning rate $\alpha$ can also be selected to control how much influence the previous estimates have on the current batch. We also leave online training of BNN using quantum annealing as crucial future work.



\section{Conclusions}

We are proposing a method to formulate Binary Neural Networks (BNNs) training  in a form able to run on latest iteration of quantum annealer hardware.
We show the promising scaling of the run time, running our algorithm on the actual hardware.
It is worth of notice that often the slowest part for running the experiments was solving the problems using classical algorithms.
The tools available to ``compile'' an algorithm into a quantum architecture are not as  developed as with classical computers yet.
Implementing an algorithm has analogies to writing machine code for a classical computer.
However, like with classical computers one does not require to know the detailed physical behaviour of every component.
Although the size of the models discussed here is small, one can envision similar methods as subroutines to train parts of a full scale deep neural network. 


\medskip

{
\small

\bibliographystyle{plain}
\bibliography{bibliography.bib}

%
%
}


\

\

\

\

\appendix

\section{Appendix}

\subsection{Penalties in matrix form}

For completeness, we show the quadratic terms used to enforce the element wise multiplications and the nonlinearities in the BNN in matrix form ${\bf q}^T {\bf Q q}$.
We can expand the equation that enforces

We first consider the element wise multiplications between weights and input features (Eq.~\ref{perceptron} of the main text).
As shown in section~\ref{sec:bnnqubo}, this multiplication is equivalent to an XNOR relation between the bit variables
$q_3 = \mathsf{XNOR}(q_1,q_2)$
.
The XNOR relation can be enforced by a quadratic penalty 
(Eq.~\ref{eq:g} of the main text). When expanded in matrix form this term becomes:
\begin{align*}
g(q_1,q_2,q_3,b)=  
\begin{bmatrix}  
q_1 &
q_2 &
q_3 &
b
\end{bmatrix}
\begin{bmatrix}
-P & 2P(1-2P) & 2P  & 8P^2 \\
0  & -P       & 2P & 8P^2 \\
0  & 0        & -P  & -4P \\
0  & 0        & 0   & -2P^2 
\end{bmatrix} 
\begin{bmatrix}  
q_1\\
q_2\\
q_3\\
b
\end{bmatrix}.
\end{align*}
The matrix at the center of this term has be added to the corresponding minor of the quadratic form ${\bf Q}$.

We now consider the terms due to the nonlinearity of the sum.
In the case of a neuron with three inputs (Eq.~\ref{eq:h} in the main text) can be rewritten as:
\begin{align*}
 h(
a ,
q^{_{(out)}},
q_0^{_{(in)}},
q_1^{_{(in)}},
q_2^{_{(in)}}   
 )= \begin{bmatrix}  
a &
q^{_{(out)}}&
q_0^{_{(in)}}&
q_1^{_{(in)}}&
q_2^{_{(in)}}   
\end{bmatrix}
\begin{bmatrix}
P & 2P & -P & -P & -P\\
0 & 4P & -2P & -2P & -2P\\
0 & 0 & P & P & P\\
0 & 0 & 0 & P & P\\
0 & 0 & 0 & 0 & P
\end{bmatrix}
\begin{bmatrix}  
a\\
q^{_{(out)}}\\
q_0^{_{(in)}}\\
q_1^{_{(in)}}\\
q_2^{_{(in)}}   
\end{bmatrix}.
\end{align*}
The terms of the square matrix need to be added to the corresponding minor of $\bf Q$ to create the quadratic loss ${\bf q}^T{\bf Q q}$.

\subsection{QUBO on a dataset}
\label{sec:appendix_dataset}

Here we show how to set up the optimization of the weights of a binary neural network as a QUBO problem in the case of a dataset of size $D$.
The total $0-1$ indicator loss on the dataset $\{({\bf y}_d^0, \hat{y}_d)\}_{d=1}^D$ of the binary neural network described in the main text can be written as:
\[L = \sum_{d=1}^D L_{0-1}(\hat{y}_d, y^L_d)\]
where $y_d^L = f_{BNN}({\bf w})({\bf y}_d^0)$, the BNN with weigths $\bf w$.
To write the QUBO formulation of this equation is enough to notice that the weights are ``shared'' for all the data.
On the other hand, instead of each one of the other binary variables we now need $D$ novel binary variables, that we index by $d$.
These binary variables represent the features 
($q_{y_{i,d}^l}$), the outputs ($q_{y_{d}^L}$),
the element-wise products
($q_{Z_{i,j,d}^{l}}$), and the ancilla variables ($a_{i,j,d}^l$ and $b_{i,j,d}^l$).

The final QUBO formulation is analogous to Eq.~\ref{eq:qubo}:

\begin{align}\label{eq:qubo_data}
\begin{aligned}
\min_{ \bf q \in { \{0,1\}^M }} \quad & \sum_{d=1}^{D} \left[ \left({ q_{y^L_d} -   \hat{b}_d}\right)^2 + \sum_{\ell,i,j} g( q_{Z^{\ell+1}_{i,j,d}}, q_{W^\ell_{i,j}}, q_{y^\ell_{i,d}}, b^\ell_{i,j,d})  + \sum_{\ell,i,j} k( a_{i,j,d}^\ell, ..., q_{y_{j,d}^{\ell+1}}, q_{Z_{k,j,d}^\ell} )\right],  \\
\end{aligned}
\end{align}
where $\hat{b}_d=(\hat{y}_d+1)/2$ is the ground truth in binary of the datum $d$.
Note that the weights do not have an index $d$ and they are shared over all the $D$ data.

\subsection{Graph visualization of QUBO}

\begin{figure}
  \centering
  \includegraphics[width=1.0\textwidth]{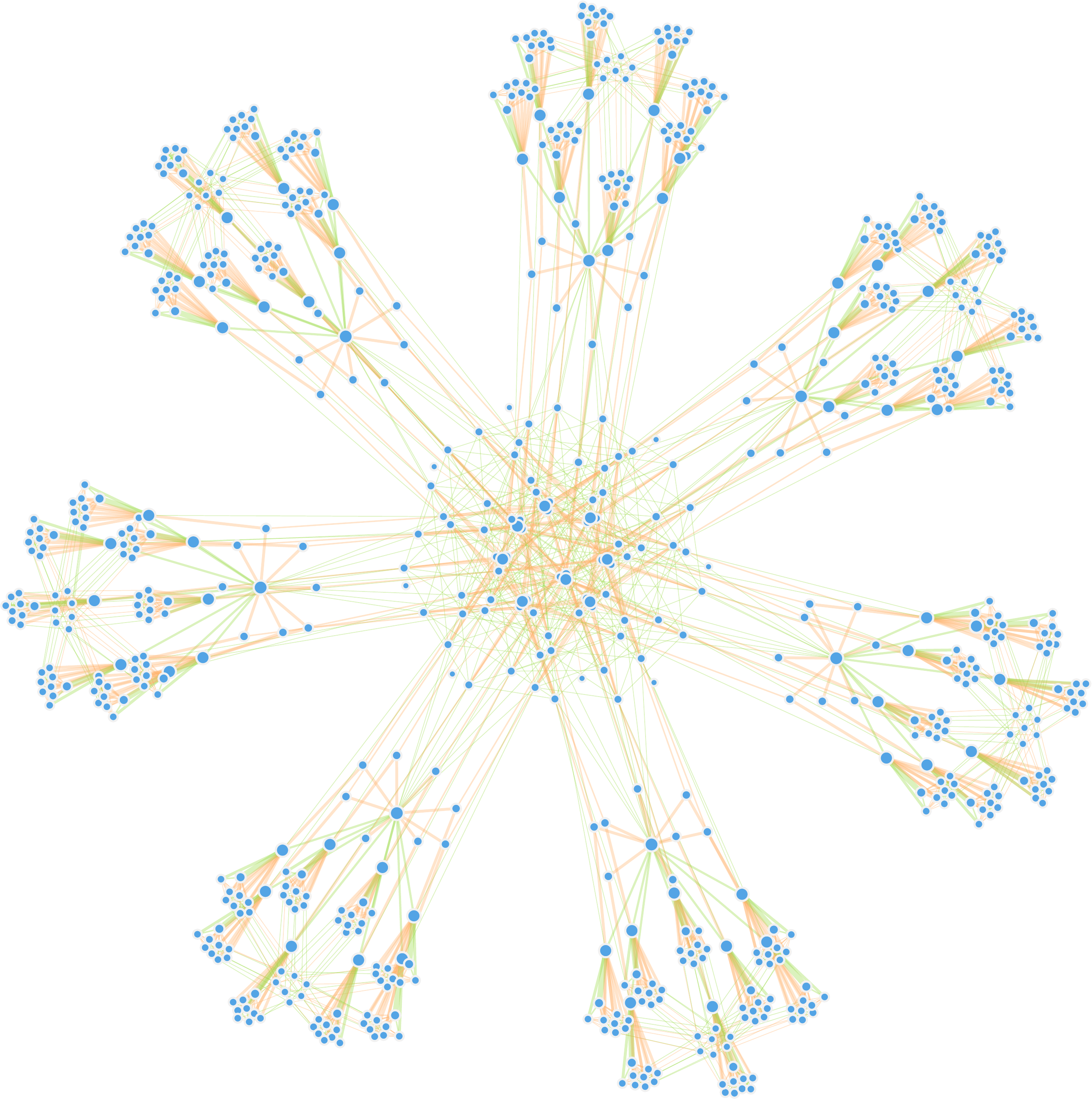}
  \caption{A graph representing the QUBO formulation of a network with $7$ inputs, $7$ neurons in one hidden layer, dataset size of $8$.}
  \label{fig:graph_qubo}
\end{figure}

We present a graph visualization for a neural network with $7$ inputs, $7$ neurons, one hidden layer and dataset size of $8$. The corresponding QUBO formulation, obtained following section~\ref{sec:bnnqubo} and~\ref{sec:appendix_dataset}, is shown in Fig.~\ref{fig:graph_qubo} using the {\it networkx}\footnote{\url{https://networkx.org/}} visualization tool.
The nodes (${\bf q}_{n}$) represent the weights, and for each datum hidden features, outputs and ancilla bits.
The edges (${\bf Q}_{nm}$) represent the the structure of the network built with penalties to connect these quantities.
The edges structure also represents the input values, the ground truth classifications and the loss function as described in section~\ref{sec:bnnqubo}.
The color and length of the edges if Fig.~\ref{fig:graph_qubo} represent respectively the sign and the strength of the ${\bf Q}_{nm}$ coefficients.
As explained in section~\ref{sec:embedding}, this formulation is (typically) not a subgraph of the QPU graph. We need to embed our QUBO in the QPU graph, following the procedure from~\cite{cai2014practical}. This final step returns the blue graph shown in Fig.~\ref{fig:embedding} in the main text.

\end{document}